# Improvement of Sensitivity of Capacitive Micromachined Ultrasound Transducer

## Yifan Wang and Yunfei Liu

Y. Wang is a Master's Graduate Student in the College of Engineering, University of Illinois at Urbana-Champaign, IL, USA (*Corresponding author e-mail: ywang123@illinois.edu).

*Abstract*— **Capacitive Micromachined Ultrasonic Transducer (CMUT) has a wild range of applications in medical detecting and imaging fields. However, operating under self-generating-self-receiving (SGSR) method usually results in poor sensitivity. But the sensitivity cannot be improved simply by increasing the resonant frequency since the frequency of a specific kind of CMUT is designed for specific usage. In this paper, based on one specific type of CMUT, mechanical model is built and simulation analysis is demonstrated. A brand-new method one-generating-multiple-receiving (OGMR) is introduced and a special circuit model has been designed to improve the signal to thermal noise ratio. By increasing the number of receiving capacitors from 1 to 8, we increased the signal-noise ratio to 2.83 times.**

## I. INTRODUCTION

Recent years have been witnessing the booming of CMUTs, Capacitive Micromachined Ultrasonic Transducers, which enable products to be smaller, more integrated, and cost-effective, accomplishing applications ranging from industrial handheld non-destructive testing, medical imaging, diagnostics to automation.

One typical CMUT we chose during the project is shown in Figure 1:

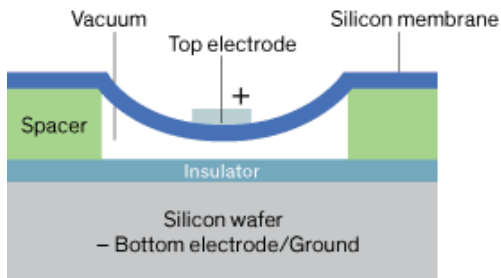

Fig. 1. CMUT Structure

A parallel plate capacitor is formed by a top metalized membrane and a bottom substrate. The membrane material is typically silicon, while doped silicon plays as the substrate in most case. As often the case, we fix the substrate up and let the membrane drift according to any internal signal. When a voltage signal is applied on the either sides of the capacitor, the membrane will deflect in the direction of the substrate due to the attractive electrostatic forces regardless of the polarity. As soon as applying a mixture of DC and AC signal, the AC signal can be regarded to have static bias. During the transmission period, the static bias will exert a slight resistive force in the direction opposite to the membrane's deflection, which resulting in the fluctuating and vibration of the membrane, thus, generating an ultrasound. The ultrasound is typically with a frequency of 2-10 MHz in bio-MEMS application. In reception turn, the reflected ultrasound exerts pressure on the membrane, so that an AC signal can be perceived.

The CMUT generally works under such self-generating-self-receiving (SGSR) operation mechanism. Due to the SGSR mechanism, CMUT is easy to fabricate, and can be manufactured in diverse sizes and shapes. However, from practical experience, the signal perceived is very weak and small, which reduced the precision and sensitivity of the sensing process. But, CMUT is designed with specific frequency aiming at specific usage, which means simply increasing the resonant frequency is not an effective way to improve the sensitivity.

In this paper, a new method named one-generating-multiple-receiving (OGMR) is achieved and a special circuit has been designed to improve the sensitivity.

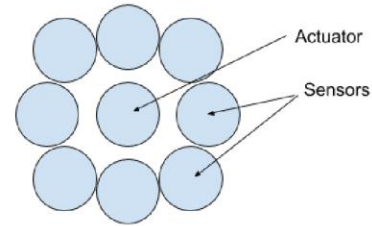

Fig. 2. One-Generating-Multiple-Receiving (OGMR) Transducer

As shown in fig 2, the basic idea of the new design is to use multiple identical sensors around the actuator to "capture" more signal information of the reflected ultrasound beam. We will discuss this new design in detail in the following sections.

## II. MODEL AND SIMULATION

### A. Mechanical Model

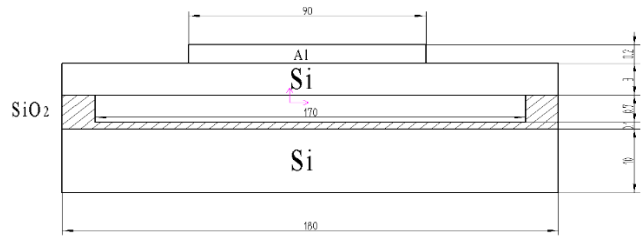

Fig. 3. Mechanical Structure of CMUT

The mechanical structure of CMUT is designed as figure 3. The upper silicon layer works as the membrane which moves vertically, producing ultrasound waves. The silicon dioxide acts as the insulator, keeping upper silicon from directly contacting the silicon substrate. We apply doping method to implant As atoms into silicon substrate to make it a N-type layer, which contains enough free electrons. Coupled with an aluminum layer, the silicon substrate is connected by a voltage signal source, both serving as electrode plates.

The model's parameters are choosen from a product. Based on those numbers, the natural frequency is calculated as:

$$f_1 = \frac{2.54 h}{\pi r_a^2} \sqrt{\frac{E}{3\rho(1-\sigma^2)}} = 1.73 \times 10^6 Hz$$

In order to simplify the computing process and make the model useful as well as near real conditions, we made some assumptions.

1. Assume membrane has fixed edge, so vibrating radius is 85 μm;

Even though the whole length of silicon membrane is 180um, the 10um fixed part is deleted for it cannot move.

2. Assume a parallel capacitor: upper plate of capacitor is upper silicon membrane; lower plate is lower silicon substrate.

In terms of the equation to calculate capacitor, there is a variable "dielectric constant", if we do not simplify the model, setting aluminum layer as upper plate of capacitor, the dielectric constant of silicon, vacuum and silicon dioxide are all contained in calculation, which makes modeling process complicated.

By assuming the condition, the simplified model is near real condition, and we can easily get the capacitor is 0.287pF.

3. Assume membrane vibration model can be simulated by a "spring-damping-mass" model, which (Mass) dose piston motion:

For circuits design part, we assume the capacitor plates are parallel to each other through vertical vibration, which brings convenience for circuits simulation. This requires us to transfer the original model (deflection on membrane's surface is not parallel to lower plate) into a new "spring-damping-mass" model, which is shown below:

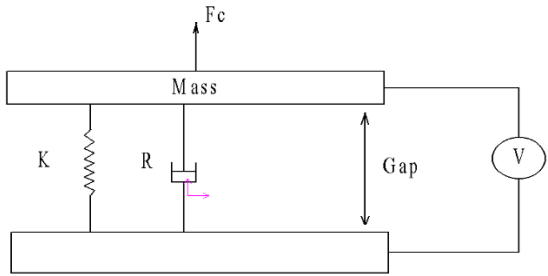

Fig. 4. Spring-Damping-Mass Model

For this model, we assume the membrane (upper silicon plate) does piston motion to simulate vibration.

4. Assume an extra-mass (Mr) into membrane mass Mm and an extra-damping (Rr) into mechanical damping Rm are added into the model;
5. Assume there are only three acting forces: spring force, damping force and electrostatic force.

These two assumptions make our new model reasonable and close to real condition. The atoms in the air near membrane have reaction to it, whose influence equals to adding a virtual mass Mr and a virtual damping Rr into corresponding variables. Besides the forces of buffer parts in model, the only existing outer force we must consider is electrostatic force Fc.

To estimate the deflection of membrane, an input electrostatic force Fc is needed, also for circuits design.

In terms of model vibration equation:

$$M\frac{d^2 w}{dt^2} + R\frac{dw}{dt} + kw = F$$

The total mass of moving part attached to membrane is:

$$M = M_m + M_r + M_{le}$$

$$M_r \approx \frac{8\rho_0 r a^3}{3}$$

$$M_{le} = \rho \mathcal{V}$$

Where Mm is the mass of upper silicon plate; Mr is adding virtual mass; Mele is the mass of aluminum layer.

The spring constant is:

$$K = \frac{192\pi D}{r_a^2} = \frac{192\pi E h^3}{r_a^2 \cdot 12(1-r^2)}$$

Where E is Young's modules and r is Poisson ratio.

The natural frequency is:

$$f_o = \frac{1}{2\pi} \cdot \sqrt{\frac{K}{M}}$$

The result is 1.753MHz. This value is very close to the original model's, which means we can use this model to simulate the deflection condition of the former mechanical structure.

The total damping is:

$$R = R_m + R_r \approx 50 \cdot R_r = 50 \times \frac{\rho_0 C_0 K^2 \pi r_a^4}{2}$$

Where Rr is adding virtual damping and C0 is the sound velocity in air.

The quality factor is:

$$Q = \frac{w_0 M_m}{R}$$

The pull-in voltage of the model is:

$$V = \sqrt{\frac{8kd^3}{27\varepsilon_o A}}$$

We set the input impulse voltage signal is the same as the pull-in voltage, enabling the membrane to generate the largest deflection. The input electrostatic force F is:

$$F = \frac{1}{2}V^2 \frac{\varepsilon_o A}{(d-x)^2}$$

And the largest deflection is:

$$w = \frac{F}{K}$$

We can figure out the voltage is 132.75V, the electrostatic force is 0.0081N and the deflection is 0.233µm.

### B. Simulation Analysis

#### 1) Model Design

Based on the CMUT model we picked, a 3D model is built and imported to FEA software. The software used in this part is Solidworks and Ansys. According to the materials, we define Silicon and Aluminum in the model. In order to make thing easier, we assume here Silicon to be isotropic elastic material with Young's modules exclusively along its <110> direction. The material properties are shown below:

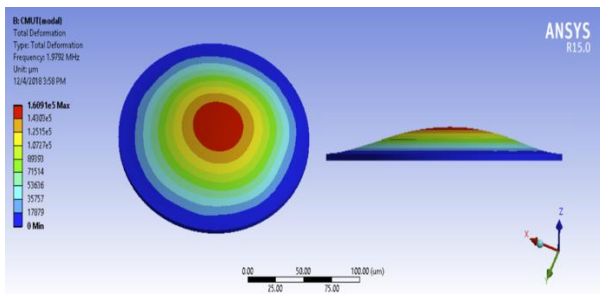

Fig. 5. Material Properties

#### 2) Modal Analysis

We first demonstrate modal analysis. As shown in Fig 6, we define the outer cylindrical surface as fixed support which aligns to the general idea of the model:

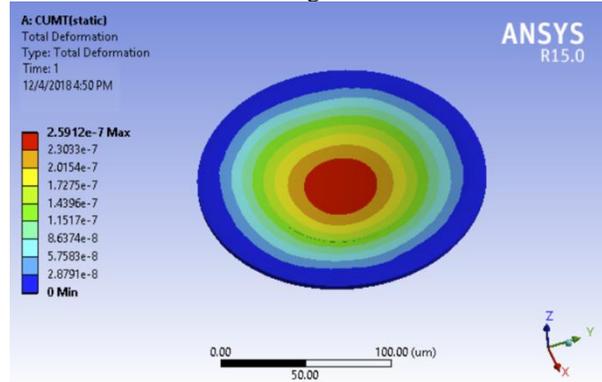

Fig. 6. Modal Simulation Result

In order to make things more intuitive, a side view is put together with the original output. The simulation result shows that the resonant frequency is 1.9792MHz, which is slightly higher than the theoretical value 1.753MHz in last section. However, in the simulation model, the pretension has been ignored and we also eliminated the mass of substrate which contributes to such reasonable upwards partial error.

#### 3) Static Analysis

Besides, from the previous derivation, at "pull-in" point, the electrical force in the capacitor is about 0.0081N. The same force has been exerted on the center of the top membrane in simulation model, verifying the theoretical deflections. The simulation result is shown in Fig 7:

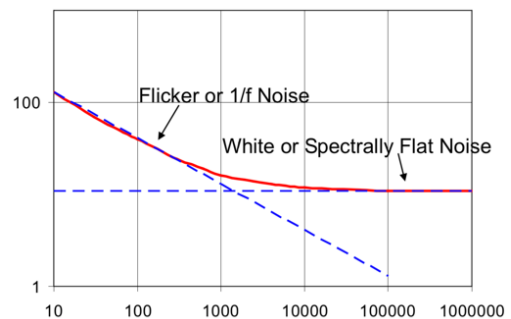

Fig. 7. Static Simulation Result

From the result we can see that the simulated deflection is 0.2591µm, which also is a little bit larger than the theoretical value 0.233µm, 1 third of the distance between two capacitor plates. Still, ignoring pretension and the material assumption account for the upwards bias.

Combining all the simulation analysis above, conclusion can be made that our model is reliable and the output is credible.

## III. CIRCUITS DESIGN

One way to increase the sensitivity is to use Iterative Learning Control (ILC) (see [11]-[13]), another way (used in this paper) is to increase the signal-noise ratio. As mentioned in previous paragraphs, the sensor works under a frequency of 1.753MHz. According to fig 8, in frequency domain, at the frequency of more than 1KHz, the white noise (thermal noise) dominates. So, in our circuit model, we will only take care of thermal noise.

Fig. 8. Noise Distribution

According to thermal noise equation, we can see that the voltage noise density is positive correlated with absolute temperature and also the total resistance in circuit:

$$\overline{V_{noise}} = \sqrt{4K_BRT}$$

This equation indicates that by putting several identical sensors in serial, the total resistance of the circuit will increase n times, and the signal voltage will also increase n times, but the voltage noise level will only increase by √n times. Based on an assumption that all these identical sensors will generate the same output signal at the same time when receiving same input, we can see that the signal-noise ratio get increased by √n times.

### A. Non-linear Solution

Figure 9 shows a typical microphone output circuit model. It is widely used in industry. The output voltage changes with the capacitance value of the sensor. The signal output of this typical circuit is:

$$V_{out} = \frac{1}{j\omega R(C+\Delta C)+1}V_{in}$$

Here ΔC is the capacitance change of the sensor. It is obvious that the output signal is non-linear with the capacitance change or the displacement of membrane of the sensor.

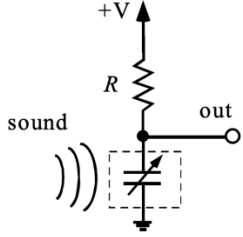

Fig. 9. Typical Circuit

Also, in this typical circuit, the noise level is:

$$V_{noise} = \sqrt{4K_BRT\Delta f + \frac{K_BT}{C}}$$

We can see that there are two separate parts in this equation. The first part refers to the thermal noise generated by the resistor R, second part is the thermal noise generated by the capacitor.

Now, to verify the assumption mentioned at first, as shown in figure 10, we put n capacitance sensors in serial to see what will happen.

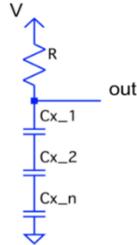

Fig. 10. Improved Sensing Circuit

The output signal is:

$$V_{out} = \frac{n}{j\omega R(C+\Delta C)+n}V_{in}$$

Here C is the capacitance of each sensor and ΔC is the capacitance change of each sensor. The thermal noise level is:

$$V_{noise} = \sqrt[2]{4K_BRT\Delta f + \frac{nK_BT}{C}}$$

We can see that the resistor part stays the same while the capacitor part increased by n times. So, we can consider the thermal noise as a non-linear function of sensor number, Vnoise(n), and also consider the output voltage as a function of sensor number and capacitance change Vout(n, ΔC). So that we can establish an equation for signal-noise ratio:

$$\eta(n) = \frac{V_{out}(n,0) - V_{out}(n,\Delta C)}{V_{noise}(n)}$$

Since we only care about how the number of sensors affects the signal-noise ratio, we can consider ΔC as a constant value. Now we substitute signal voltage and noise level into signal-noise ratio, we have:

$$\eta(n) = \frac{\left[\frac{nR\frac{\Delta C}{C}}{(\frac{C+\Delta C}{C}R+nZ)(R+nZ)}\right]V_{in}}{\sqrt[2]{4K_BRT\Delta f + \frac{nK_BT}{C}}}$$

$$Z = \frac{1}{j\omega C}$$

$$\Delta f = \frac{n}{2\pi RC}$$

Where Z is the impedance of each capacitor, Δf is the bandwidth of the whole circuit. To simplify this equation, we make an approximation that (C+ΔC)/C≈1, since ΔC<<C. The simplified signal-noise ratio function is:

$$\eta(n) = K\frac{\sqrt{n}}{(1+n\frac{Z}{R})^2}$$

$$K = \frac{\frac{\Delta C}{C}V_{in}}{\sqrt[2]{\frac{K_BT}{C}(\frac{2}{\pi}+1)}}$$

For a certain circuit, K is constant. From the equations above, we can see it is obvious that the signal-noise ratio is negatively correlated with sensor number n. However, if we make the resistance of R is a function of n, R(n)=nZ, which means the resistor value changes with the number of capacitors in the circuit, we will have an increasing function:

$$\eta(n) = K\frac{\sqrt{n}}{4}$$

Assume we have maximum displacement Δx=d/3 (d/3 is the maxim displacement of actuator part. Generally speaking, the reflected ultrasound beam will cause a deflection less than s/3). The plot of function η(n)=1.29e4√n is shown in figure 11:

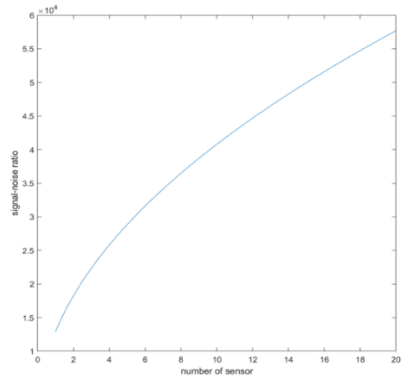

Fig. 11. Signal-Noise Ratio Increased with n

## B. Linear Solution

Although the signal-noise ratio is increased in last section, a non-linear solution in engineering field is still not ideal. That's why we will try the linear approach to see how the model will perform. Fig 12 shows a typical circuit that usually used to linearize the output of capacitor sensor:

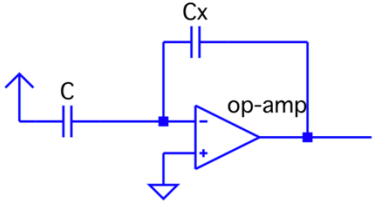

Fig. 12. Typical Linearizing Circuit

We can see that it is an inverting configured op-amplifier. We chose OPA695 to build our circuit. Z is the impedance if capacitor. The output signal of this circuit is:

$$V_{out} = \left(-\frac{Z_x}{Z_0}\right) V_{in}$$

$$C_x = \frac{\varepsilon A}{d-x}, \quad C_0 = \frac{\varepsilon A}{d}$$

$$V_{out} = -V_{in} + \frac{x}{d} V_{in}$$

The output voltage is separated into two parts, (-Vin) is the DC offset, and the rest is the output AC signal. By adding a RC filter at the output part of the circuit, we can easily remove the DC offset.

Similarly, based on the assumption mentioned at the start of circuit design section, we put sensors in serial like fig 13:

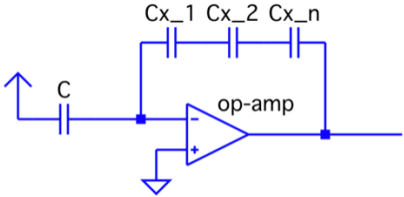

Fig. 13. Imporved Linear Circuit

In such circuit, the gain of the op-amplifier is G=-n. Obviously, both the DC offset part and signal part get amplified to -n times:

$$V_{out} = \left(-\frac{nZ_x}{Z_0}\right) V_{in}$$

$$V_{out} = -nV_{in} + \frac{nx}{d} V_{in}$$

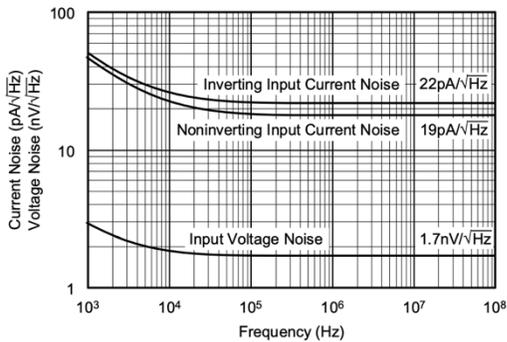

Fig. 14. Input Current &Voltage Noise

According to the datasheet of OPA695 shown in fig 14, the inverting input current noise density and the input voltage noise density are:

$$\overline{I_N} = 22pA/\sqrt{Hz}, \quad \overline{V_N} = 19nV/\sqrt{Hz}$$

$$V_{noise} = \sqrt{\frac{K_B T}{\frac{n+1}{n} C_0}(n+1) + \left(\frac{n}{n+1}\overline{I_N}Z\right)^2 \Delta f(n+1) + \overline{V_N}^2 \Delta f n}$$

We can see that there are three parts of non-correlated noise in this circuit. The first part is thermal noise generated by capacitor sensors and C0; the second part is the voltage noise that transferred from current noise; the third part is the voltage noise generated by the amplifier itself.

Similarly, we use η to represent the signal-noise ratio:

$$\eta(n) = \frac{V_{signal}(n)}{V_{noise}(n)} = \frac{n\frac{x}{d}V_{in}}{\sqrt{\frac{nK_B T}{C_0} + (\overline{I_N}Z)^2 \Delta f\left(\frac{n^2}{n+1}\right) + \overline{V_N}^2 \Delta f n}}$$

Substitute the actual value into η(n), we have：

$$\eta(n) = \frac{1.67\sqrt{n}}{\sqrt{1.44\times10^{-8} + 4.83\times10^{-11}\Delta f\frac{n}{n+1} + 2.8\times10^{-18}\Delta f}}$$

We found that in terms of thermal noise in this function, the first part is significantly higher than the rest of two. So, if we generated by op-amp itself, we will have a function as shown in fig15:

$$\eta(n) = \frac{1.67\sqrt{n}}{\sqrt{1.44\times10^{-8}}} = 1.39\times10^4\sqrt{n}$$

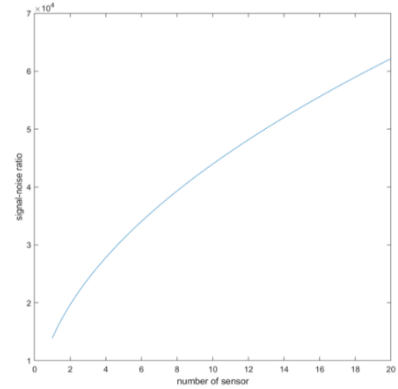

Fig. 15. Signal-Noise Ratio Increased with n

However, if we take the current noise and voltage noise into consideration, things can be more complicated. Since the voltage noise and current noise are both correlated with bandwidth Δf. In terms of bandwidth, we assume that it is also a function of n. But unfortunately, we don't know how Δf will change when n changes.

## IV. FABRICATION

The fabrication of our device can be done from surface micromachining. During the process, we grow several layers of different materials and lift off useless parts in a particular sequence. Photolithography technique is considered and masks are carefully designed. Among these are three important steps, first, the vacuum cave under the upper silicon layer is anisotropically etched. This procedure provides enough space for the deflection of membrane vibration. Then a thin film of aluminum of specific shape is grown, attached with upper silicon layer, serving as upper electrode. Finally, some vertical sidewalls are etched to settle the lower electrode to wire package.

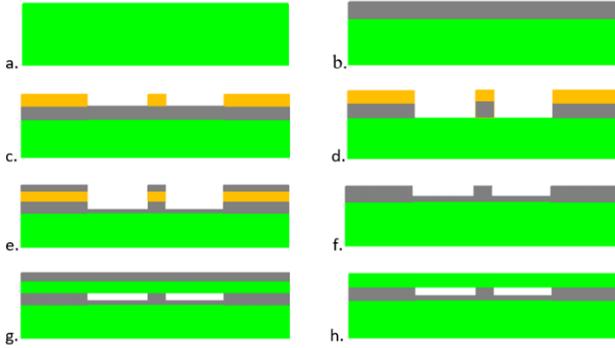

Fig. 16. Model Fabrication Process Flow. a)ion implantation, making 4"N-type <100> silicon. b) thermal oxidation. c) photolithography. d) plasma etching. e) thermal oxidation. f) lift off. g) wafer bonding. h) lift off.

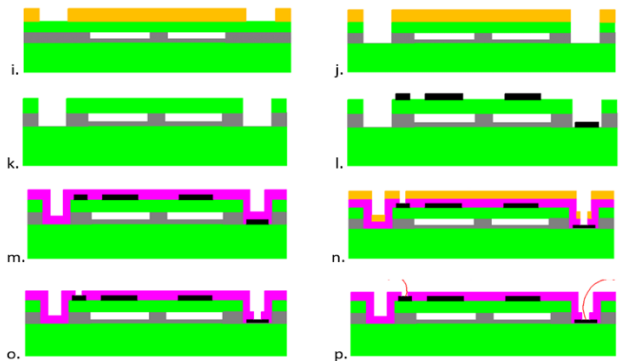

Fig. 17. Model Fabrication Process Flow. i) photolithography. j) ion milling & plasma etching. k) strip photoresist. l) evaporate aluminum & photolithography. m) sputter LTO. n) photolithography & plasma etching. o) strip photoresist. p) wire bond & test.

| | Si | | PR | | Al |
|---|---|---|---|---|---|
| | SiO₂ | | LTO | | Cu |

Fig. 18. Materials of Layers

## V. Result and Conclusion

Two solutions for increasing the sensitivity of capacitive ultrasound transducer are verified in this paper. By splitting the actuating and sensing circuit apart and increasing the number of sensors. In terms of the non-linear solution, a number of n identical sensors are connected in serial. By increasing signal output and noise at the same time but in different way, we successfully increased the signal-noise ratio by √n times. As for the linear solution, the signal to external-input-noise ratio is increased, but the ratio to internal noise is still unclear because of unknowing of the relationship between Δf and sensor number n.

Further research may focus on figuring out the relationship between bandwidth and sensor number to see whether this linear solution actually works.